\documentclass[useAMS,usenatbib]{mn2e}

\usepackage{graphicx}

\title[Episodic Explosions in Interstellar Ices]
{Episodic Explosions in Interstellar Ices}

\author[J.M.C. Rawlings et al.]
{J.M.C.~Rawlings,$^1$\thanks{E--mail: jcr@star.ucl.ac.uk}
D.A.~Williams, $^1$
S.~Viti,$^1$
C.~Cecchi\textendash Pestellini,$^2$ 
\newauthor 
and W.W.~Duley,$^3$ \\
$^1$University College London, Department of Physics and Astronomy,  
Gower Street, London WC1E 6BT, United Kingdom\\
$^2$INAF ~\textendash~ Osservatorio Astronomico di Cagliari, Strada n.54, 
Loc. Poggio dei Pini, 09012 Capoterra (CA), Italy\\
$^3$University of Waterloo, Department of Physics and Astronomy, Waterloo,
Ontario, Canada N2L 3G1}

\begin{document}

\maketitle

\begin{abstract}
We present a model for the formation of large organic molecules in dark 
clouds. The molecules are produced in the high density gas-phase that exists
immediately after ice mantles are explosively sublimated. The explosions are 
initiated by the catastrophic recombination of trapped atomic hydrogen.

We propose that, in molecular clouds, the processes of freeze-out onto ice
mantles, accumulation of radicals, explosion and then rapid (three-body) 
gas-phase chemistry occurs in a cyclic fashion. This can lead to a cumulative
molecular enrichment of the interstellar medium.

A model of the time-dependent chemistries, based on this hypothesis, shows that
significant abundances of large molecular species can be formed, although the 
complexity of the species is limited by the short expansion timescale
in the gas, immediately following mantle explosion.

We find that this mechanism may be an important source of smaller organic 
species, such as methanol and formaldehyde, as well as precursors to 
bio-molecule formation.
Most significantly, we predict the gas-phase presence of these larger molecular 
species in quiescent molecular clouds and not just dynamically active regions, 
such as hot cores.
As such the mechanism that we propose complements alternative methods of large 
molecule formation, such as those that invoke solid-state chemistry within 
activated ice mantles. 

\end{abstract}


\section{Introduction}

A variety of laboratory evidence shows that the catastrophic recombination of 
hydrogen atoms and other accumulated radicals in a solid may abruptly 
raise the temperature of the solid to $\sim 10^3$ K. \citet{DW11} proposed that
for HAC dust grains in the interstellar medium, such temperatures are
sufficiently high to permit the grains to radiate in the so-called Unidentified 
Infrared Bands (UIBs) associated with carbonaceous materials, lying 
between 3.3 and 11.3 $\mu$m. Recently, \citet{CPDW12} 
have shown that H$_2$ molecules released from HAC grains in such abrupt 
temperature excursions caused by the recombination of a sufficient number 
of accumulated H-atoms could not only account for the populations in 
rotational levels of H$_2$ observed in the ``hot component'' of H$_2$ in 
diffuse clouds, but also produce amounts of CH$^+$ and OH comparable with 
those observed in diffuse clouds.
This theory of radical recombination-driven mantle explosions is very similar 
to that described by \citet{G76} except that the mechanism for mantle 
explosion is the spontaneous internal recombination of trapped hydrogen atoms, 
rather than an external heating source. 

In this paper, we consider an application of the idea of abrupt 
temperature excursions caused by radical recombination to dark clouds where 
ices have been deposited on dust grain surfaces. If the abrupt 
temperature excursions are of the magnitude discussed by \citet{DW11}
then any ices present will be abruptly converted to gas and evaporated 
explosively. As pointed out by \citet{CCP10}, hereafter Paper I,
the density in the expanding gas can be so high that three-body 
reactions have the potential to create new and more complex species from 
the ice constituents. Here, we extend the ideas of \citet{CCP10} to the gas 
formed by the exploding ices.

Interstellar ices are deposited on dust grains in dark interstellar clouds 
in regions where the visual extinction exceeds some critical value, 
typically a few visual magnitudes. The ices in quiescent lines of sight 
towards low mass stars are observed to contain H$_2$O, CO, CO$_2$, CH$_4$, 
NH$_3$, H$_2$CO, CH$_3$OH, OCN$^-$, and some other species of low abundance 
\citep{OBP11}. The ices may also contain radicals created from these molecules 
by photodissociation, where the dissociating radiation is the cosmic 
ray-induced radiation field \citep{PT83}.

Thus, H$_2$O may give rise to OH, CH$_4$ to 
CH$_3$, etc., and the population of these radicals is associated with the 
bulk of the ice. 
The various reactions resulting from the accretion of atoms and 
molecules from the ambient gas as well as processing by ultraviolet photons and
cosmic rays must always be incomplete and will, for example, lead to the
presence of trapped H-atoms at weak binding sites on or near the surface.
According to \citet{DW11}, when a critical number of these weakly bound 
H-atoms is accumulated, 
then a localised recombination of a few hydrogen atoms can trigger 
a chemical runaway which releases all of the chemical energy stored in the 
grain. This can include both the hydrogen recombination energy and the 
energy stored in other radicals as well as the grain substrate.
This leads to a runaway explosion which abruptly heats the ice mantle and 
grain core to temperatures on the order of one thousand K. \citet{DW11} 
show that the number of H-atoms required to cause this explosion 
is equivalent to about 5\% of the total number of atoms in the grain plus 
mantle. We postulate that the recombination of trapped H atoms to H$_2$ 
drives the explosion but that the heavy, and less mobile, molecular 
radicals are released into the gas phase. 

In the work presented here, we follow the time-dependent gas-phase 
chemistry in a dark interstellar cloud; we compute the deposition of ices 
and the accumulation of weakly bound H-atoms that are assumed to trigger 
the explosions when their number on a grain reaches the critical value. 
Thus, in this model, the frequency of the episodic temperature excursions 
is controlled by the cloud chemistry. When a temperature excursion 
occurs, the ices are assumed to be instantaneously sublimated, and 
three-body reactions involving radicals produce products in the very high 
density expanding gas. This gas is assumed to undergo a free expansion 
which occurs on a timescale on the order of nanoseconds; however, the 
density in this sublimate is so high that many collisions occur before the 
gas has relaxed to more normal interstellar conditions. In this picture, 
the episodic explosions therefore enrich the interstellar gas with the 
products of the three-body reactions, and the material undergoing this 
enrichment is accumulated during the interval between explosions. The 
interstellar gas may be enriched by a number of successive events, if the 
timescales for accumulation of ices and H-atoms (typically, on the order 
of one million years in canonical dark cloud conditions) are suitable. 
Evidently, if the interval is very short (as will occur if the H-atom 
abundance in the gas is high), little material can be accumulated as ices. 

Our model for molecular cloud chemistry has some similarity to that
described by \citet{GWWH08} to account for chemical complexity in hot
cores. The accumulation of radicals within ices is adopted in both our model
of chemistry in molecular clouds and that of Garrod et al. in hot cores. In
our model, the radicals are released through the abrupt warming in episodic
explosions, while in the Garrod et al. model of hot core chemistry the 
radicals are released in the slow warm-up of material in the early hot core 
phase.

We describe the model in Section 2. The results are presented in Section 
3, and we give a discussion of the implications and make our conclusions 
in Section 4.

\section{The model}

To test the hypothesis described above, we first of all constructed a simple 
`proof of concept' model of the time-dependent chemistry. The first stage 
considers the chemistry (gas-phase, plus freeze-out), for moderately dense
conditions (density $n=10^3$cm$^{-3}$, temperature $T=10$\,K, extinction $A_V=3$)
where we expect the atomic hydrogen abundance to be high, so as promote the
rapid accumulation of radicals in the ice mantles.
During this phase limited surface chemistry is allowed to occur, so that full 
hydrogenation of C, N, O, and S-hydrides to CH$_4$, NH$_3$, H$_2$O and H$_2$S
are allowed. A fraction (10\%) of the CO is allowed to react with O and OH 
on the surface of grains to form CO$_2$. 
It is further assumed that no (continuous) desorption mechanisms are operating.
After a specified period ($\sim 10^4-10^6$ years) the ice mantles (composed of
CO, CH$_4$, NH$_3$, N$_2$, O$_2$, H$_2$O, Na, H$_2$CO, CO$_2$, HCN, HNC, HNO, 
H$_2$S, C$_2$S, HCS, O and OH) are released back into the gas phase. 
No change of temperature or density is included and no high-density chemistry 
is included. The cycle is then allowed to repeat.

The results from this model show that, for most species, a limit cycle is 
achieved after only one or two mantle explosions. Ions and unsaturated species 
tend not to be strongly affected by the process, but the simple saturated 
species (such as H$_2$O, CH$_4$ etc.) are very strongly enhanced. 
A few species, such as CO and CO$_2$ show a slower, more steady rise in 
abundance over several cycles.

We have therefore constructed a more realistic and comprehensive two-phase 
model which considers the time-dependent chemistry at a single point in a 
molecular cloud.
The parameters for this model are given in Table~\ref{tab:parameters}.
This model utilizes the LSODE integration package \citep{HP95} and has been 
sub-divided to describe the two chemical phases. 
Phase I represents the (standard) dark cloud chemistry, with freeze-out and 
(limited) surface chemistry, whilst Phase II considers the chemistry in the 
high density, rapidly expanding gas, in the immediate vicinity of a dust grain 
following ice mantle sublimation.

The two phases are physically and chemically distinct from each other, but the 
output from each phase feeds into the other as material cycles between the two.
The characteristics of the two phases are described below:-

\subsection{Phase I}

The physical and chemical parameters for Phase I (as given in 
Table~\ref{tab:parameters}) are discussed below.

\begin{table}
\caption[models]{Physical parameters in the standard model - see text for 
description}
\begin{tabular}{|l|c|}
\hline
Parameter  & Value \\ 
\hline
 He/H & 0.1 \\
 C/H & $2.6\times 10^{-4}$ \\
 N/H & $6.1\times 10^{-5}$ \\
 O/H & $4.6\times 10^{-4}$ \\
 S/H & $1.0\times 10^{-7}$ \\
 Na/H & $1.0\times 10^{-7}$ \\
\hline
 Density ($n_I$) & $10^4$ cm$^{-3}$ \\
 Temperature ($T_I$) & 10 K \\
 $A_V$ &  3 magnitudes \\
 Cosmic ray ionization rate ($\zeta_0$) & $1.3\times 10^{-17}$ s$^{-1}$ \\
 Initial abundance of H-atoms ($n_{H,0}$) & 1 cm$^{-3}$ (see text) \\
\hline
 H-atom non-recombination probability ($p_H$) & 0.1 \\
 Explosion threshold abundance of H ($f_H$) & 0.05 \\
 No. of (refractory) atoms per grain ($N_g$) & $10^8$ \\
 Mantle radical formation rate ($R_{rad}$) & 0.01 Myr$^{-1}$ \\
\hline
 Average grain radius ($a$) & 0.0083 $\mu$m \\
 Dust surface area per H-nucleon ($\sigma_H$) & 
 $8.0\times 10^{-21}$ cm$^2$ \\
 Grain albedo & 0.5 \\ 
 CO $\to$ CO$_2$ conversion efficiency ($f_{CO_2}$) & 0.1 \\
\hline
 Phase II: Initial density ($n_{II}$) & $10^{20}$ cm$^{-3}$ \\
 Phase II: Initial Temperature & 1000 K \\
 Phase II: Three-body rate coefficients ($k_{3B}$) & 
   $10^{-28}$ cm$^6$s$^{-1}$ \\
\hline
 Number of cycles ($n_{cyc.}$) & 5 \\
\hline
\end{tabular}
\label{tab:parameters}
\end{table}

For the gas-grain interactions,
all species are assumed to have a sticking coefficient of 1.0 and
as with the simple model - and for the purpose of clarity - we suppress all 
continuous desorption processes. The freeze-out rates (and hence the H to H$_2$
conversion rate) are calculated self-consistently, using the mean grain radius 
($a$) and surface area per hydrogen nucleon ($\sigma_H$) as specified in 
Table~\ref{tab:parameters}.
These values are consistent with standard values for the dust-to-gas ratio
and freeze-out rates in dark clouds \citep[e.g.][]{RHMW92}.
We use a typical value for the cosmic ray ionization rate and the reaction
network is drawn from the UMIST06 database \citep{W07}.
We also include photodissociations by the secondary radiation field generated 
by the cosmic ray ionization and excitation of H$_2$ \citep{PT83} which
have an inverse dependence on $\sigma_H$ and $(1-\omega)$, where $\omega$ is 
the average grain albedo.

We assume that full hydration of atoms and simple hydrides to CH$_4$, NH$_3$, 
H$_2$O and H$_2$S occurs on grain surfaces.
As in other studies \citep[e.g.][]{RK12} we assume that a fraction 
($f_{CO_2}$) of the CO that impacts a grain and interacts with surface O or OH
is converted into CO$_2$. Some models of hot cores \citep[e.g.][]{VW99} invoke 
the partial conversion of CO to
CH$_3$OH and/or H$_2$CO on the surface of grains to explain the high gas-phase 
abundances that are observed. Whether or not such processes are efficient 
\citep{WK02} we have not included them in our model as we wish to distinguish 
the role of the proposed mechanism in the generation of larger organic species.

The extinction ($A_V$) is taken to be 3 magnitudes for $n_I=10^4$cm$^{-3}$ 
and 10 for higher densities.
It is assumed that the chemistry is `dark' - i.e. well outside any 
photon-dominated regions (PDRs), so that the photodissociation rates for H$_2$
and CO and the photoionization rate for C are all set to zero. 

During this phase we assume that reactive radicals are created in the ice
mantles due to the action of impinging cosmic rays. We note that there may be 
contributions both from direct cosmic ray impact and photolysis by the cosmic 
ray induced radiation field \citep{PT83}.
This is a cumulative effect and we assume that it is limited to the stripping 
of a single hydrogen atom from saturated species. Thus:
\[ {\rm CH_4 \to CH_3 + H} \]
\[ {\rm NH_3 \to NH_2 + H} \]
etc. This applies to; CH$_4$, NH$_3$, H$_2$O, OH, H$_2$CO, H$_2$S, CH$_3$OH,
HCN, HNC, HNO and HCS. The relative abundance of the radical to the parent
saturated species will therefore be proportional to the cosmic ray ionization 
rate and the period of exposure, $t$. This fraction is then given by:-
\[ \frac{n_{rad}}{n_{sat}} = R_{rad}\left(\frac{t}{1~Myr^{-1}}\right)
\left(\frac{\zeta_0}{1.3\times 10^{-17}s^{-1}}\right) \]
Note that we have not included the (equally likely) possibility of the 
formation of molecular ions in the ice mantles. This is a potentially serious 
omission, but since we have little idea as to the efficiencies and products we 
have opted not to include these processes in our model. However, we also note
that molecular ions are highly reactive, so the results from this study must 
be regarded as a lower limit to the formation efficiencies of complex organic 
molecules.

It is further assumed that the hydrogen atoms so-generated are released back
into the gas-phase and are {\em not} retained in the ice mantles.
These radicals will be crucial in determining the efficiency of large molecule
formation in Phase II, as we postulate that radical-radical three-body 
reactions will be very much faster (by a factor of $>100\times$) than reactions
between saturated species. 
We use a value of $R_{rad}=1$\% per Myear in our standard model.

In the case of atomic hydrogen we assume that a fraction ($p_H$) that 
impacts grains simply sticks as free hydrogen atoms in the ice. The remainder
recombines to H$_2$ and is returned to the gas phase, as per standard models
of interstellar clouds. In our standard model we follow \citet{DW11} and adopt
a value of 0.1 for $p_H$.
The timescale between mantle explosions is determined by the rate of accretion
of atomic hydrogen, so the initial H abundance is an important parameter.
Assuming that chemical equilibrium initially pertains in the cloud then, for
typical gas-to-dust ratios, the abundance of atomic hydrogen is given by 
\citep[e.g.][]{DW84}
\[ n_H = 1\left(\frac{\zeta_0}{1.3\times 10^{-17}{\rm s}^{-1}} 
\right) {\rm cm}^{-3} \]

There are 81 gas-phase and 25 solid-state species in the dark cloud (Phase I) 
chemistry which are listed in 
Table~\ref{tab:species1}. Those given in italics are solid-state (frozen-out)
species, whilst those in bold are the solid-state radicals formed 
{\em in situ}. 

\begin{table}
\caption[models]{Chemical species in Phase I}
\begin{tabular}{|c|}
\hline
H, H$_2$, H$^+$, H$^-$, H$_2^+$, H$_3^+$, He, He$^+$, Na, Na$^+$, \\
C, C$^+$, C$^-$, CO, CO$^+$, CH, CH$^+$, CH$_2$, CH$_2^+$, \\
CH$_3$, CH$_3^+$, CH$_4$, CH$_4^+$, CH$_5^+$, \\
N, N$^+$, NH, NH$^+$, NH$_2$, NH$_2^+$, NH$_3$, NH$_3^+$, NH$_4^+$, N$_2$, N$_2^+$, N$_2$H$^+$, \\
O, O$^+$, O$_2$, O$_2^+$, OH, OH$^+$, H$_2$O, H$_2$O$^+$, H$_3$O$^+$, \\
HCO, HCO$^+$, H$_2$CO, H$_2$CO$^+$, CO$_2$, CO$_2^+$, HCO$_2^+$, \\
CN, CN$^+$, HCN, HCN$^+$, HNC, \\
NO, NO$^+$, HNO, HNO$^+$, HCNH$^+$, H$_2$NC$^+$, HNCO$^+$, H$_2$NO$^+$, \\
S, S$^+$, HS, HS$^+$, H$_2$S, H$_2$S$^+$, H$_3$S$^+$, \\
CS, CS$^+$, C$_2$S, C$_2$S$^+$, HC$_2$S$^+$, HCS, HCS$^+$, H$_2$CS$^+$, \\
{\em CH$_4$, CO, NH$_3$, N$_2$, O$_2$, H$_2$O, Na, H$_2$CO, CO$_2$, HCN, HNC,} \\
{\em HNO, H$_2$S, C$_2$S, HCS, O, OH, H, CH$_3$OH,} \\
{\bf CH$_3$, NH$_2$, HCO, HS, CH$_3$O, CH$_2$OH} \\
\hline
\end{tabular}
\label{tab:species1}
\end{table}

The time-dependence of the chemistry is followed until the atomic hydrogen 
abundance in the ices reaches some threshold value ($f_H$). At this point we 
assume that the hydrogen explosively recombines (with 100\% efficiency) to 
H$_2$ and all components of the ice mantles are instantaneously heated and 
fully sublimated. 
The energy liberated has to heat the whole grain, so $f_H$ is the fraction
of H-atoms relative to the ice mantle, plus the refractory core.
Thus the threshold abundance of H is given by:
\[  \frac{n_g(H)}{N_g+\sum_i n_g(i)} = f_H \]
where $n_g(H)$ is the number of hydrogen atoms in the ice mantle,
$N_g$ is the (average) number of refractory atoms/molecules per grain 
and $\sum_i n_g(i)$ is the sum of all the atoms/molecules in the ice mantle.
\citet{DW11} showed that to raise the grain temperature to $\sim$1000\,K, 
12 kJ/mole of stored energy is required, as verified by experiment.
Obviously, much less energy would 
be needed if the required rise in temperature were smaller (e.g. 100\,K would
probably be enough to initiate radical recombination in ice mantles). 
This quantity defines the required concentration of reactive species and so,
taking the conservative value of 12 kJ/mole, in our standard model we follow
\citet{DW11} and adopt a value of 0.05 for $f_H$.

\subsection{Phase II}

There are 34 chemical species in the Phase (II) high-density expanding gas as 
listed in Table~\ref{tab:species2}. The (limited) reaction network is 
shown in Table~\ref{tab:network2}. 
Both the species and the reaction network are highly uncertain and 
ill-determined. We speculate that the three-body reactions (for which we 
assume the third body to be H$_2$O, the most abundant species in the 
post-sublimation gas-phase) may yield obvious organics (such as 
C$_2$H$_5$OH and (CH$_2$OH)$_2$).
Of course other, unidentified, reaction channels/products may also exist. 
Other (small species) products of the reactions are not specified in 
Table~\ref{tab:network2}.
Since clear detections of sulfur-bearing species in ice mantles do not exist, 
we do not include any complex species/reactions which incorporate sulfur 
atoms.
As H$_2$O is the dominant gas-phase constituent of the Phase II gas, we 
renormalise all integrations to calculate abundances relative to H$_2$O, 
rather than H nucleons.

\begin{table}
\caption[models]{Chemical species in Phase II}
\begin{tabular}{|c|}
\hline
H$_2$O, H, H$_2$, CO, CH$_4$, CH$_3$, NH$_3$, NH$_2$, OH, H$_2$CO, HCO, \\ 
CO$_2$, H$_2$S, HS, CH$_3$OH, CH$_3$O, CH$_2$OH, NH$_2$OH, HCOOH, \\
C$_2$H$_6$, CH$_3$NH$_2$, CH$_3$CH$_3$O, C$_2$H$_5$OH, CH$_3$CHO, HCONH$_2$, \\
(CH$_2$OH)$_2$, CH$_2$OHCHO, CH$_2$OHNH$_2$, CH$_3$OCH$_3$O, \\
CH$_3$OH, CH$_3$OCH$_2$OH \\
\hline
\end{tabular}
\label{tab:species2}
\end{table}

\begin{table}
%\begin{minipage}{180mm}
\caption{Reaction network in Phase II}
\label{tab:network2}
    \begin{tabular}{ccccccc}
\\
 Reaction & & & & & & \\
  \hline
  OH      & + & CH$_3$  & + & H$_2$O & $\to$ & CH$_3$OH \\ 
  OH      & + & NH$_2$  & + & H$_2$O  & $\to$ & NH$_2$OH \\
  OH      & + & CH$_3$O & + & H$_2$O & $\to$ & CH$_3$OOH \\
  OH      & + & HCO  & + & H$_2$O & $\to$ & HCOOH \\
  CH$_3$  & + & CH$_3$     & + & H$_2$O & $\to$ & C$_2$H$_6$ \\
  CH$_3$  & + & NH$_2$ & + & H$_2$O & $\to$ & CH$_3$NH$_2$ \\
  CH$_3$  & + & CH$_3$O & + & H$_2$O & $\to$ & (CH$_3$)$_2$O \\
  CH$_3$  & + & CH$_2$OH  & + & H$_2$O & $\to$ & C$_2$H$_5$OH \\ 
  CH$_3$  & + & HCO  & + & H$_2$O & $\to$ & CH$_3$CHO \\ 
  NH$_2$  & + & CH$_2$OH & + & H$_2$O & $\to$ & CH$_2$OHNH$_2$ \\ 
  NH$_2$  & + & HCO & + & H$_2$O & $\to$ & HCONH$_2$ \\ 
  CH$_3$O & + & CH$_3$O & + & H$_2$O & $\to$ & CH$_3$OCH$_3$O \\ 
  CH$_3$O & + & CH$_2$OH & + & H$_2$O & $\to$ & CH$_3$OCH$_2$OH \\ 
  CH$_2$OH & + & CH$_2$OH & + & H$_2$O & $\to$ & (CH$_2$OH)$_2$ \\ 
  CH$_2$OH & + & HCO & + & H$_2$O & $\to$ & CH$_2$OHCHO \\ 
    \hline
    \end{tabular}
%  \end{minipage}
\end{table}

The physical conditions in this extremely high density, rapidly expanding, gas
are as described in Paper I, so that we again consider a sphere of ice, 
instantaneously sublimated into the gas-phase of initial radius $r_0$ and 
with an initial density ($n_0$) that is comparable to the solid-state number
density. This gas freely expands, with spherical symmetry, into a vacuum at 
the sound speed $v_s$.
Then, as described in Paper I, if $r_0$ is assumed to be comparable to the 
typical thickness of an ice mantle and $v=10^4$~cm\,s$^{-1}$, the  
evolution of the number density is given by
\begin{equation}
n/n_0 = 1 / \left( 1+10^{9} t \right)^3 
\end{equation}
where $t$ is the time in seconds.
At such high densities the chemistry will be completely dominated by three-body
reactions.
Some of the mantle explosion energy will be released as heat and we assume that
the gas has an initial temperature of $\sim 1000$\,K, which may help to drive 
the chemistry. 
We follow the practice of Paper I and adopt a single value for the rate 
coefficient (k$_{3B}$) for all reactions which incorporates any implicit 
dependence on temperature. The radical-radical-H$_2$O reactions are expected
to be faster than the radical-neutral-H$_2$O reactions and so we adopt larger 
values for k$_{3B}$ than we did in Paper I.
As we do not include any specific temperature-dependences of the rates, 
the variation of temperature with time is not a relevant parameter in our
models.

We follow the chemistry in this expanding sublimated gas - which is quite 
distinct from the background chemistry in the surrounding (Phase I) cloud - 
until quasi-equilibrium and/or suppression 
of chemical activity due to geometrical expansion occurs. In practice, this is
extremely fast - of the order of 1-10\,ns (see Paper I); many orders of 
magnitude faster than the chemical timescale of Phase I. This necessitates a 
separate numerical integration and the complete switching of the chemistry and 
differential equations at each transition between Phases I and II. 

\subsection{Cycling}

After 10\,ns, when the Phase II chemical calculations are terminated,
the abundances (calculated relative to H$_2$O) are 
re-normalised to hydrogen and (fully) mixed back into the Phase I gas.
The cycle, as described above and the duration of which is calculated 
self-consistently, is then repeated.
For the standard model ($\zeta_0=1.3\times 10 ^{-17}$s$^{-1}$) the 
Phase I duration is $\sim 3.2\times 10^5$ years and we follow the 
chemical evolution through 5 cycles. 
We use this number of cycles for two reasons: (i)
The results indicate that, by then, the solution is approaching a limit cycle, and (ii) The total duration of $\sim 1.5$\,Myr is comparable to the 
average lifetime for a dense cloud. However, for 
$\zeta_0=1.3\times 10 ^{-16}$\,s$^{-1}$, $n_{\rm H}\sim 10$\,cm$^{-3}$ and so the 
inter-explosion period is reduced accordingly to $\sim 3.2-4\times 10^4$ years. 
We therefore run the calculation for 50 cycles, so as cover approximately the
same time interval ($\sim 1.5$\,Myr) as for the standard model. 

There will obviously be some optimal value for $n_H$; if it is too high, then 
that implies that the molecular content of the gas is low and ices will not 
form. If it is too low, then the inter-explosion timescale becomes too long
for the processes described here to be important.
The importance of $n_H$ in controlling mantle growth in molecular clouds may
be relevant to the determination of the observed critical visual extinction
for the onset of ice formation. We shall return to this question in a later
publication.

Strictly speaking, we should model the gas-phase chemistry and freeze-out of 
the larger organic molecules created in Phase II through successive cycles 
to ever increasing complexity.
However, we note that the chemistry is very much faster and more efficient in 
Phase II, so we make the simplifying assumption that the larger species (not 
included in the species list for Phase I) are `sinked' and simply build up in
successive cycles, not taking part in the gas-phase chemistry, and not being
converted into larger species. 
By contrast, simple saturated species (such as CH$_4$ and NH$_3$) are allowed 
to re-cycle with a full gas-phase chemistry, freeze-out and sublimation.

\section{Results}

At first sight it may seem that there are an unmanageably large number of free
parameters in the model, but in reality many of these are degenerate and in
effect the chemical complexity will largely be defined by two ratios:
\begin{enumerate}
\item In Phase I; the ratio of the accretion to the inter-explosion timescales, 
and
\item In Phase II; the ratio of the three-body chemistry to the expansion
timescales
\end{enumerate}
In the discussion below,
we therefore only investigate the sensitivities to four parameters; the density
in Phase I ($n_I$), the cosmic ray ionization rate ($\zeta_0$), the density in 
Phase II ($n_{II}$) and the rate of formation of radicals in the ice mantle
($R_{rad}$).
Other parameters, such as the assumed universal three-body reaction rate 
($k_{3B}$), have obvious linear effects on the large species formation 
efficiencies (subject to saturation limits).

The model described above is complex; yielding abundances as a function of time
for both the quiescent phase (Phase I) and the sublimation phase (Phase II) for 
each of the cycles of quiescence and explosion.
We present the results in two forms: (i) as graphical representations of the 
time-dependence of (key) species in the two phases - results from all cycles 
being shown on one plot, and (ii) as tabulated abundances of species in 
Phase I.

\begin{figure*}
\includegraphics[scale=0.75]{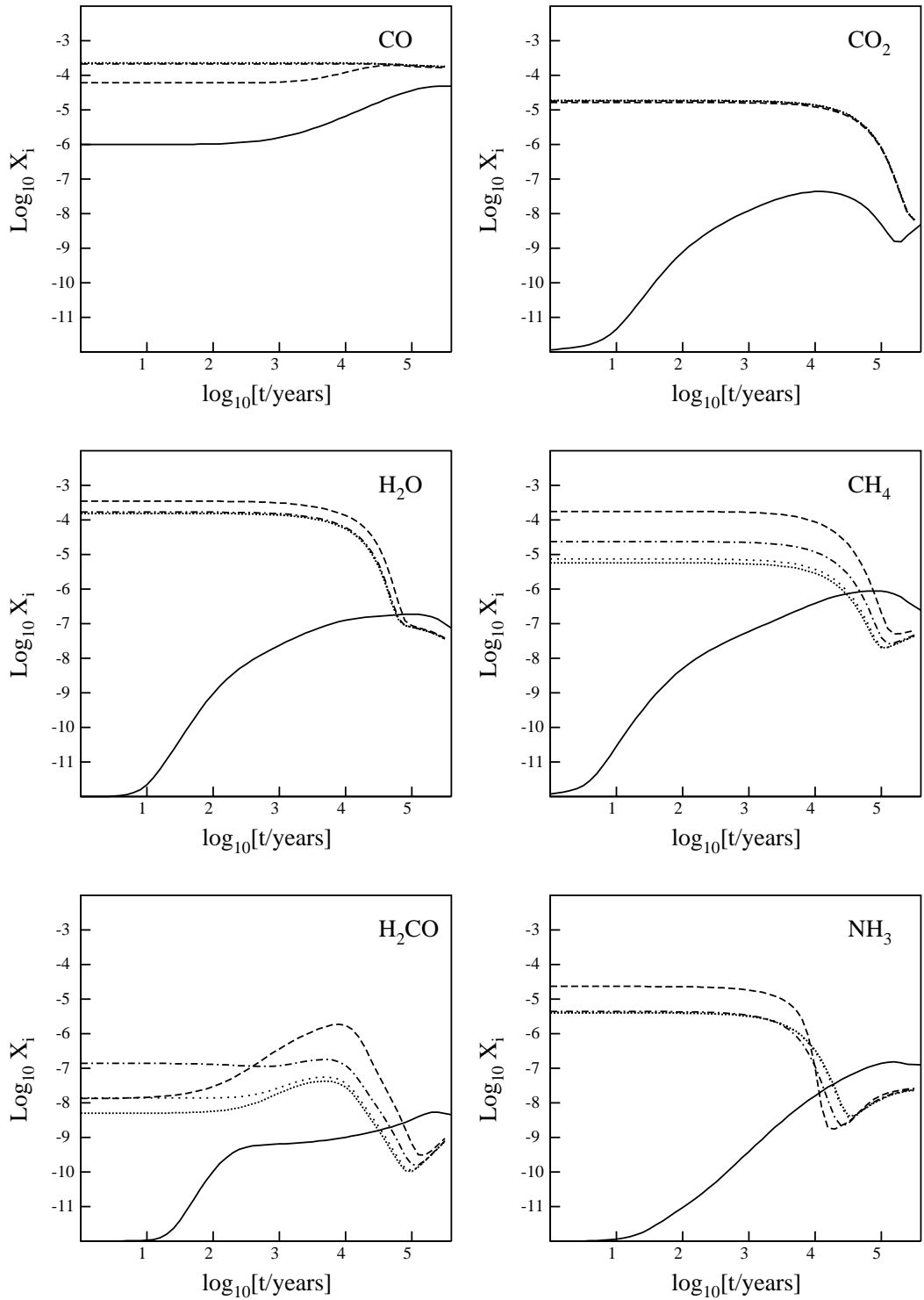}
\caption{Logarithmic fractional abundances (relative to H-nucleons) of 
selected species as a function of time in Phase I.
Results are shown, overlaid, for each of the five cycles: 1-solid line, 
2-dashed line, 3-dot-dashed line, 4-wide dotted line, 5-close-dotted line.}
\label{fig:phaseI}
\end{figure*}

\begin{figure*}
\includegraphics[scale=0.75]{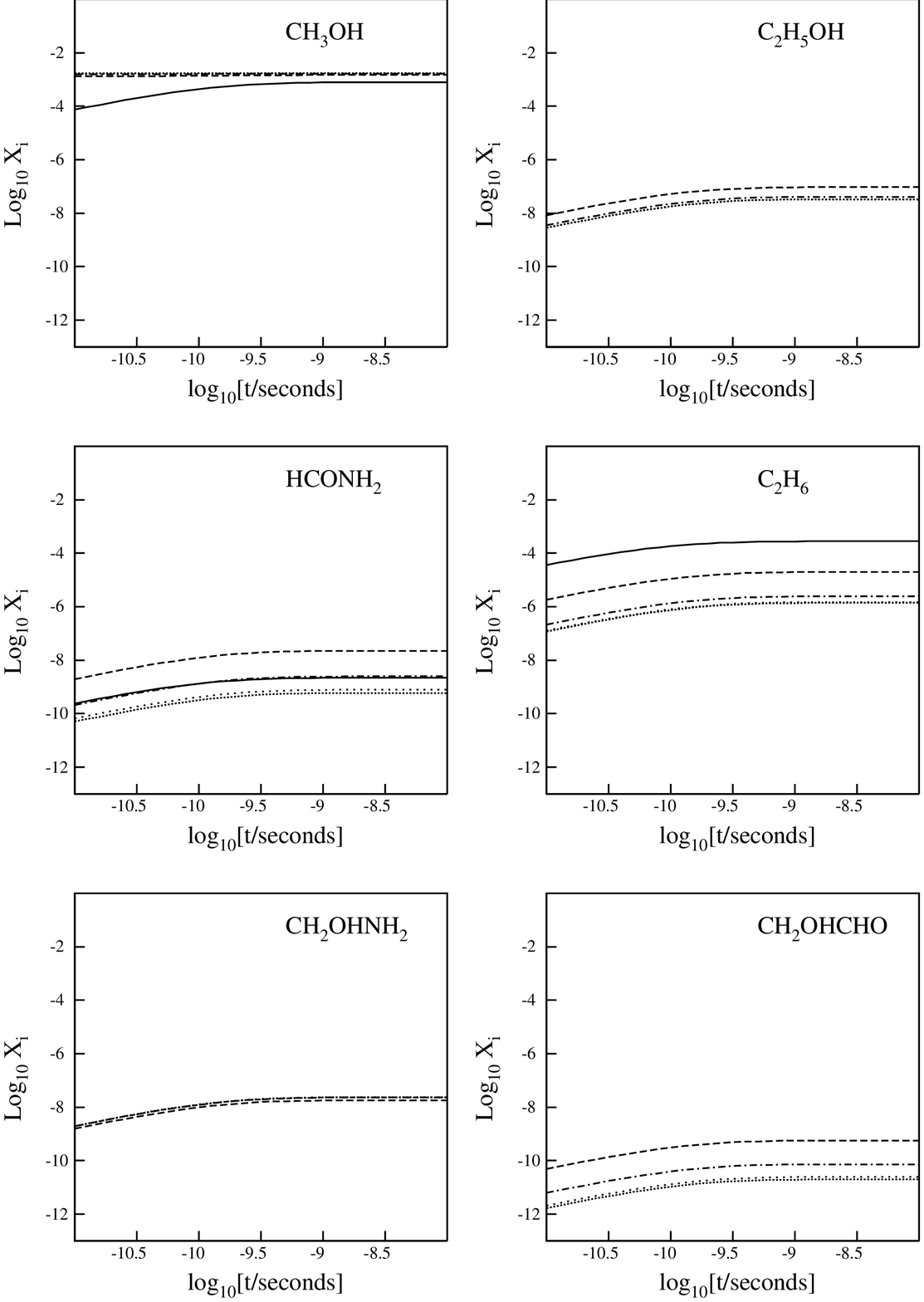}
\caption{Logarithmic fractional abundances (relative to H$_2$O) of selected 
species as a function of time in Phase II.
Results are shown, overlaid, for each of the five cycles:1-solid line, 
2-dashed line, 3-dot-dashed line, 4-wide-dotted line, 5-close-dotted line.}
\label{fig:phaseII}
\end{figure*}

Figures~\ref{fig:phaseI} and \ref{fig:phaseII} show the abundances of selected
species in Phases I and II respectively.
The different curves in each frame depict the results from different cycles.
The time-dependences in Phase II, shown in Fig~\ref{fig:phaseII}, are simple 
and as expected; molecular abundances build up very rapidly until saturation
(exhaustion of the reactants) and/or geometrical dilution become important
(typically within 1\,ns).
From Fig~\ref{fig:phaseI} it is immediately apparent that, in Phase I, there is 
considerable variation of the abundances with time, partly due to the gas-phase 
chemistry, but mainly as a result of the freeze-out of species onto dust 
grains. Interstellar clouds will consist of an incoherent
ensemble of material in various stages of cycling between the gas-phase and 
the frozen-out states.
Bearing these two facts in mind it is therefore apparent that to make a 
sensible comparison with observations we need to calculate the 
{\em time-averaged} values of the abundances. 
For those species where we calculate the full time-dependence of the 
abundances the time-averaged values can be easily obtained. For the larger
species produced in Phase II, whose gas-phase chemistry we do not follow,
we use a simple analytical approximation.

On the asumption that freeze-out is the {\em only} process that affects the 
abundances of these species in Phase I, the rate of change of abundance of
species $i$ is simply;
\[ \dot{n}_i = -k_in_i.n \]
where $k_i$ is the freeze-out rate for species $i$ and $n$ is the density.
Simple integration then yields, for the average fractional abundance;
\[ \langle x_i\rangle = \frac{X_i(0)}{k_i n t} \left[ 1-e^{-k_i n t} 
\right] \]
where $X_i(0)$ is the fractional abundance of species $i$ at the beginning 
of Phase I and $t$ is the duration of the phase.
This approach therefore makes allowance for the statistical ensemble of
explosion cycles, but does not include the effects of sequential enrichment 
of the gas through successive explosions. Whilst some species will approach a 
limit cycle, others will be constantly supplied by successive explosions and
a cumulative build-up of the explosion products will occur.
To quantify these effects we have studied the (time-averaged) abundances in a 
sequence of cycles - see Table~~\ref{tab:results1} below.

We also note that other destruction processes may need to be considered. 
In the case of the larger organic molecules, protonation followed by 
dissociative recombination may be an effective loss mechanism 
\citep[e.g.][]{BRW06}.
We can make a simple, empirically-based, estimate for the efficiency of 
this process as follows: Assume physical parameters as for our 
standard model and a fractional ionization of $\sim 10^{-8}-10^{-7}$, as 
appropriate for a molecular cloud. As an upper limit, if we further assume 
that the abundance of the protonating reactant is equal to the ionization 
fraction and that the reaction rate coefficient is 
$\sim 10^{-9}$\,cm$^3$s$^{-1}$ then the upper limit to the protonation (and 
loss) rate implies a destruction timescale of 
$\sim 3\times 10^4 - 3\times 10^5$\,years.
To check this approximation we have run a simple model of a static 
molecular cloud with a full gas-phase chemistry. The initial abundance of
CH$_3$OH is set high ($\sim 10^{-7}$). The model confirms that the CH$_3$OH
decays on these timescales.  

These values are comparable to the time interval between explosions.
By comparison, in our standard model, the freeze-out timescale for CH$_3$OH 
is $\sim 2\times 10^5$\,years. 
We therefore conclude that chemical destruction of the larger molecules is
possibly significant, inhibiting the (time-averaged) abundances by a factor 
of $\sim 2-5$, although this effect would be reduced significantly if the 
period betweeen explosions were shorter.

We present the results from 10 model runs. Model 1 is the standard model, 
which uses the parameter values given in Table~\ref{tab:parameters}.
The variations to these parameter values used in the other models are given 
in Table~\ref{tab:models}.

Tables~\ref{tab:results1}, \ref{tab:results2} and \ref{tab:results3}
give the (Phase I) time-averaged abundances for selected species calculated 
as described above.
Note that the only species in these tables that are {\em not} created 
exclusively in the explosions are NH$_3$, H$_2$O, H$_2$CO and H$_2$S.
These are all formed in both the gas and the solid-state.
In the absence of explosions, all other species would have a zero gas-phase
abundance and, since we suppress continuous desorption in our models, the 
smaller species would also just freeze-out within 1\,Myr.
Table~\ref{tab:results1} shows results for the standard model (model 1) in 
the form of the (average) abundances of selected species in each of the five 
cycles.
Several conclusions can be drawn from this table: The small (Phase I) 
gas-phase species approach a limit cycle - with only small cycle-to-cycle 
variations after 4 or 5 cycles. The same applies to the smaller species 
produced in the explosion chemistry (eg. CH$_3$OH, C$_2$H$_6$) which reach 
abundances of $\sim 10^{-9}-10^{-7}$. The larger species (eg. CH$_2$OHNH$_2$,
C$_2$H$_5$OH) have smaller abundances ($\sim 10^{-13}-10^{-10}$) but these 
steadily grow as each cycle adds more to the `reservoir' of the larger species.
  
Tables~\ref{tab:results2} and \ref{tab:results3} show the sensitivity of the 
results to variations in the free parameters.
The results from models 2-4 are a little surprising at first sight; increasing 
the (Phase I) density drives a faster and more efficient chemistry, so that
the abundances are somewhat enhanced for model 2 
($n(I)=10^5$\,cm$^{-3}$). However at higher densities (in models 3 and 4), as 
might be applicable to hot core environments, the abundances are notably 
suppressed - especially for the larger species; model 4 yields the lowest
abundances of any of the models. This is simply a result of the rapid 
freeze-out at high densities so that, although the abundances at the 
beginnning of Phase I are higher, the time-averaged abundances in Phase I 
are much lower. 
Increasing the cosmic ray ionization rate ($\zeta$) - model 5 - again yields 
relatively small enhancements relative to model 1 (by factors of $\leq 2-6$) 
for most species, whilst some are inhibited. However, increasing {\em both} 
$\zeta$ and $n_I$ (model 6) results in substantial abundance enhancements, 
especially for the larger species which in some cases (eg. CH$_3$OCH$_3$O) are 
by more than three orders of magnitude. 
Part of the explanation for this is that higher 
ionization rates result in larger atomic hydrogen abundances and hence shorter 
inter-explosion periods. This, in turn, means that the gas-phase depletion 
factors are lower.

Models 7 and 8 investigate the effects of a higher density in Phase II 
($n_{II}$). In model 8 the density in Phase I is also increased.
In the case of model 7 the chemistry of the Phase I species is barely affected,
whilst the species formed in the explosions are enhanced by small factors 
(typically $\leq 2$). Model 8 exhibits a similar behaviour to models 2-4 in 
that the higher Phase I density and shorter freeze-out timescales results in  
lower abundances.
Models 9 \& 10 investigate the effects of increasing the radical formation 
rate in the ice mantles ($R_{rad}$). In these cases the abundances of both of 
the radical reactants in the Phase II chemistry are enhanced so that there is 
a second-order effect. As a result the abundances of the larger species in 
model 10 are strongly enhanced, by factors of $\sim 6-400$.  

\begin{table}
%\begin{minipage}{180mm}
\caption{Parameter values used in the models.}
\label{tab:models}
    \begin{tabular}{ll}
\\
 Model & Parameter values \\
  \hline
 1 & standard \\
 2 & $n_I$=10$^5$\,cm$^{-3}$, A$_v$=10 \\
 3 & $n_I$=10$^6$\,cm$^{-3}$, A$_v$=10 \\
 4 & $n_I$=10$^7$\,cm$^{-3}$, A$_v$=10 \\
 5 & $\zeta$=1.3$\times 10^{-16}$\,s$^{-1}$, $n_{H,0}$=10\,cm$^{-3}$, 
n$_{cyc.}$=50 \\
 6 & $\zeta$=1.3$\times 10^{-16}$\,s$^{-1}$, $n_I$=10$^5$\,cm$^{-3}$, 
  $n_{H,0}$=10\,cm$^{-3}$, \\
   & ~~~~~A$_v$=10, n$_{cyc.}$=50 \\
\hline
 7 & $n_{II}$=10$^{21}$\,cm$^{-3}$ \\
 8 & $n_I$=10$^5$\,cm$^{-3}$, $n_{II}$=10$^{21}$\,cm$^{-3}$, A$_v$=10 \\ 
 9 & $R_{rad}$=0.02 Myr$^{-1}$ \\
 10 & $R_{rad}$=0.05 Myr$^{-1}$ \\
  \hline
    \end{tabular}
%  \end{minipage}
\end{table}

\begin{table*}
\begin{minipage}{110mm}
\caption{Time-averaged fractional abundances of selected species in each 
cycle for the standard model (run 1). A dash indicates a zero or negligibly 
small abundance. The nomenclature {\em a(b)} implies a value of 
$a\times 10^b$.}
\label{tab:results1}
    \begin{tabular}{llllll}
\\
 Species & Cycle 1 & Cycle 2 & Cycle 3 & Cycle 4 & Cycle 5 \\
  \hline
 NH$_3$       & 1.2(-7)  &  2.3(-7) & 5.7(-8)  & 6.3(-8)  & 6.3(-8) \\ 
 H$_2$O       & 1.3(-7)  &  1.0(-5) & 4.6(-6)  & 4.2(-6)  & 4.1(-6) \\
 H$_2$CO      & 4.1(-9)  &  7.4(-8) & 7.1(-9)  & 2.3(-9)  & 1.9(-9) \\
 H$_2$S       & 9.5(-14) & 7.4(-10) & 6.3(-10) & 6.4(-10) & 6.4(-10) \\
 CH$_3$OH     & -         &  1.4(-7) & 1.5(-7)  & 1.6(-7)  & 1.6(-7) \\
 NH$_2$OH     & -        & 2.0(-8)  & 2.4(-8)  & 2.6(-8)  & 2.9(-8) \\
 HCOOH        & -        & 8.2(-12) & 1.1(-10) & 1.2(-10) & 1.3(-10) \\
 C$_2$H$_6$   & -        & 4.7(-8)  & 4.9(-8)  & 4.9(-8)  & 4.9(-8) \\
 CH$_3$NH$_2$ & -        & 7.0(-9)  & 7.3(-9)  & 7.4(-9)  & 7.5(-9) \\
 CH$_3$CH$_3$O & -       & -        & 8.8(-12) & 1.2(-11) & 1.5(-11) \\
 C$_2$H$_5$OH & -        & -        & 8.8(-12) & 1.2(-11) & 1.5(-11) \\
 CH$_3$CHO    & -        & 2.8(-12) & 1.4(-11) & 1.4(-11) & 1.4(-11) \\
 HCONH$_2$    & -        & 4.2(-13) & 2.5(-12) & 2.7(-12) & 2.8(-12) \\
 (CH$_2$OH)$_2$ & -      & -        & 4.5(-14) & 1.1(-13) & 1.8(-13) \\
 CH$_2$OHCHO  & -        & -        & 5.6(-14) & 6.3(-14) & 6.5(-14) \\
 CH$_2$OHNH$_2$ & -      & -        & 1.7(-12) & 3.7(-12) & 5.7(-12) \\
 CH$_3$OCH$_3$O & -      & -        & 4.5(-14) & 1.1(-13) & 1.8(-13) \\
 CH$_3$OOH    & -        & -        & 7.8(-11) & 1.7(-10) & 2.6(-10) \\
 CH$_3$OCH$_2$OH & -     & -        & 3.9(-14) & 9.3(-14) & 1.5(-13) \\
  \hline
    \end{tabular}
  \end{minipage}
\end{table*}

\begin{table*}
\begin{minipage}{125mm}
\caption{Time-averaged fractional abundances of selected species in the
final cycle for models 1-6. The nomenclature {\em a(b)} implies a value 
of $a\times 10^b$.}
\label{tab:results2}
    \begin{tabular}{lllllll}
\\
 Species & Model 1 & Model 2 & Model 3 & Model 4 & Model 5 & Model 6 \\
  \hline
 NH$_3$       & 6.3(-8) & 1.9(-6) & 3.3(-7) & 3.5(-8) & 1.1(-7) & 5.0(-8) \\
 H$_2$O       & 4.1(-6) & 1.6(-5) & 1.6(-6) & 1.6(-7) & 7.3(-6) & 1.2(-5) \\
 H$_2$CO      & 1.9(-9) & 2.5(-8) & 4.5(-10) & 2.4(-11) & 1.0(-9) & 1.8(-9) \\
 H$_2$S       & 6.4(-10) & 4.3(-9) & 5.2(-10) & 5.2(-11) & 7.9(-10) & 4.3(-9) \\
 CH$_3$OH     & 1.6(-7) & 5.5(-8) & 5.5(-9) & 5.5(-10) & 3.2(-7) & 8.7(-7) \\
 NH$_2$OH     & 2.9(-8) & 1.3(-8) & 1.7(-9) & 1.7(-10) & 9.2(-8) & 3.9(-8) \\ 
 HCOOH        & 1.3(-10) & 7.7(-11) & 1.3(-12) & 6.9(-14) & 4.7(-11) & 4.7(-9) \\ 
 C$_2$H$_6$   & 4.9(-8) & 2.1(-8) & 2.1(-9) & 2.1(-10) & 9.9(-8) & 2.2(-7) \\
 CH$_3$NH$_2$ & 7.5(-9) & 4.7(-9) & 6.3(-10) & 6.5(-11) & 1.3(-8) & 1.2(-8) \\  
 CH$_3$CH$_3$O & 1.5(-11) & 2.2(-11) & 2.3(-12) & 2.2(-13) & 4.4(-11) & 2.0(-9) \\ 
 C$_2$H$_5$OH & 1.5(-11) & 2.2(-11) & 2.3(-12) & 2.2(-13) & 4.4(-11) & 2.0(-9) \\
 CH$_3$CHO    & 1.4(-11) & 2.9(-11) & 4.8(-13) & 2.6(-14) & 6.6(-12) & 1.1(-9) \\
 HCONH$_2$    & 2.8(-12) & 6.3(-12) & 1.4(-13) & 8.1(-15) & 9.1(-13) & 1.9(-11) \\ 
 (CH$_2$OH)$_2$ & 1.8(-13) & 3.6(-14) & 3.5(-15) & 3.5(-16) & 8.5(-13) & 2.1(-10) \\
 CH$_2$OHCHO  & 6.5(-14) & 4.3(-14) & 7.1(-16) & 3.7(-17) & 1.6(-14) & 6.8(-12) \\
 CH$_2$OHNH$_2$ & 5.7(-12) & 4.9(-12) & 6.7(-13) & 6.9(-14) & 3.3(-11) & 7.2(-11) \\
 CH$_3$OCH$_3$O & 1.8(-13) & 3.6(-14) & 3.5(-15) & 3.5(-16) & 8.5(-13) & 2.1(-10) \\
 CH$_3$OOH    & 2.6(-10) & 5.1(-11) & 5.1(-12) & 5.0(-13) & 1.6(-9) & 2.1(-8) \\
 CH$_3$OCH$_2$OH & 1.5(-13) & 2.7(-14) & 2.7(-15) & 2.6(-16) & 8.3(-13) & 1.8(-10) \\
  \hline
    \end{tabular}
  \end{minipage}
\end{table*}

\begin{table}
%\begin{minipage}{180mm}
\caption{Time-averaged fractional abundances of selected species in the
final cycle for models 7-10. The nomenclature {\em a(b)} implies a value 
of $a\times 10^b$.}
\label{tab:results3}
    \begin{tabular}{lllll}
\\
 Species & Model 7 & Model 8 & Model 9 & Model 10 \\
  \hline
 NH$_3$         & 6.3(-8) & 1.9(-6) & 6.2(-8) & 6.2(-8) \\
 H$_2$O       	& 4.1(-6) & 1.6(-5) & 4.1(-6) & 4.1(-6) \\ 
 H$_2$CO      	& 1.8(-9) & 1.9(-8) & 1.8(-9) & 1.8(-9) \\ 
 H$_2$S       	& 6.4(-10) & 4.3(-9) & 6.4(-10) & 6.3(-10) \\
 CH$_3$OH     	& 2.7(-7) & 1.0(-7) & 4.3(-7) & 1.3(-6) \\
 NH$_2$OH     	& 5.7(-8) & 2.7(-8) & 8.4(-8) & 2.7(-7) \\ 
 HCOOH     	& 2.6(-10) & 1.3(-10) & 3.8(-10) & 1.2(-9) \\
 C$_2$H$_6$     & 5.9(-8) & 2.6(-8) & 1.1(-7) & 2.9(-7) \\
 CH$_3$NH$_2$   & 9.3(-9) & 6.3(-9) & 1.8(-8) & 4.7(-8) \\ 
 CH$_3$CH$_3$O  & 3.5(-11) & 5.4(-11) & 1.0(-10) & 8.2(-10) \\
 C$_2$H$_5$OH   & 3.5(-11) & 5.4(-11) & 1.0(-10) & 8.2(-10) \\
 CH$_3$CHO    	& 1.9(-11) & 3.0(-11) & 3.5(-11) & 9.2(-11) \\
 HCONH$_2$      & 3.8(-12) & 7.1(-12) & 7.0(-12) & 1.9(-11) \\
 (CH$_2$OH)$_2$ & 7.0(-13) & 1.7(-13) & 3.2(-12) & 7.8(-11) \\
 CH$_2$OHCHO  	& 1.5(-13) & 8.7(-14) & 4.4(-13) & 3.5(-12) \\
 CH$_2$OHNH$_2$ & 1.3(-11) & 1.3(-11) & 3.9(-11) & 3.2(-10) \\
 CH$_3$OCH$_3$O & 7.0(-13) & 1.7(-13) & 3.2(-12) & 7.8(-11) \\
 CH$_3$OOH	& 9.3(-10) & 1.9(-10) & 2.1(-9) & 2.1(-8) \\
 CH$_3$OCH$_2$OH & 6.0(-13) & 1.3(-13) & 2.7(-12) & 6.7(-11) \\
  \hline
    \end{tabular}
%  \end{minipage}
\end{table}

\section{Discussion and Conclusions}

The results show that, despite the large number of poorly-constrained 
free parameters, the results are fairly insensitive within the range of 
values that we have investigated.
Significant enhancements of the abundances of large molecular species are 
obtained in the cases of a combination of an elevated cosmic ray ionization 
together with a high gas-phase density, or when then the rate of formation 
of radicals in the ices is increased. 

It is interesting to compare the predictions of our dark cloud model with 
observations.
However, in so doing, it is important to note that a direct comparison is not 
possible; a major conclusion of our study and result from the model is that we 
predict the presence of appreciable abundances of large molecules in dark 
clouds. So far, it has only been possible to detect these species towards hot 
core sources.
Never the less, despite the very different physical conditions, our 
models are able to reproduce the hot core abundances  for all
species except for the largest ones, e.g. acetamide (HCONH$_2$) and
glycolaldehyde (CH$_2$OHCHO): 
However we note that acetamide has only been observed towards the Galactic 
Centre whilst glycolaldehyde has been detected towards the Galactic 
Center and one hot core only, hence the observed abundances may not be 
typical.

Our model is general and not specific to any one source. However, although
we defer specific modelling and a detailed comparison with the dark cloud 
source TMC-1 (CP) to a future study, we can make some rough comparisons.
Some of the species given in these tables 
(NH$_3$, H$_2$O, H$_2$CO, H$_2$S, CH$_3$OH, HCOOH and CH$_3$CHO) have been 
observed in TMC-1 (CP) and their abundances have been
determined by \citet{SHC04}. These abundances are
generally in very good agreement with our models, with the notable 
exception of methanol (CH$_3$OH) which in most of our models is apparently 
over-produced by a factor of up to $\sim$100. 
However, there are two reasons why this is probably not discrepant; (i) there 
is considerable flexibility/uncertainty in the physical parameters (given in 
Table~\ref{tab:parameters}) which determine the efficiency of the formation 
of complex molecules, and (ii) as described in the previous section, we have
probably underestimated the loss rate for large molecules in the quiescent 
phase (Phase I). 
Our model also over-produces water. However, the clumps in TMC-1 are 
known to have substructure \citep{PLV98} which means that ultraviolet
penetration into CP is easier than it would be in a uniform region - as we have
assumed. So, we would expect that the time-averaged H$_2$O abundance should 
be lower than the values we have computed for the non-specific case. 
In addition, we also note that the efficiency of the surface reaction 
converting O to H$_2$O is required by observations to be high, but is
undetermined. We have assumed a value of 100\% but a factor of a few less than 
this would still be compatible with the observed strengths of the water ice 
features.
Looking ahead, we expect that the predicted presence of hot-core type molecules
within molecular clouds are at levels that can be detected with new facilities,
such as ALMA.

As it stands, the model is largely hypothetical. We have been
able to give reasonable estimates for the key parameters. However, we require
further experimental study of the explosion mechanism. 
In addition, we need confirmation of the viability of the proposed three-body
reaction pathways and quantification of the reaction rates. 
However, several conclusions can already be drawn from this study;

\begin{itemize}

\item
From our results it can be seen that our model 
predicts the presence of detectable (observationally verifiable) gas-phase 
abundances of  `large species' in molecular clouds.
However, the mechanism that we describe has a limiting effect on the size of 
molecules that can be made, because of the very short time available for 
rapid 3-body reactions to take place. 

\item 
Our model also suggest that there could be significant amounts of
undetectable molecules, such as C$_2$H$_6$, present in dark clouds. If so, 
these species - which tend to be ignored in chemical 
networks - may also play a role in the formation of observed larger species,
through ordinary gas phase ion-molecule reactions.
For example, a reaction sequence involving C$_2$H$_6$ and CH/C$_2$H 
radicals can lead to the formation of benzene (C$_6$H$_6$) in dark cloud
environments \citep{J11}.

\item
In recent years, spectral line surveys of dark cores, such as those
of TMC-1, have revealed a surprising repertoire of complex molecules
and large carbon chain species including propylene (CH$_2$CHCH$_3$). This 
latter species, expected in hot cores rather than in dark clouds, has been
detected in TMC-1 \citep{M07} but not in Orion. In
general, other more common species, such as HNCO and CH$_3$OH, are
suprisingly abundant in some dark cores. Pure gas-phase models at low
temperatures ($\sim$10 K) cannot reproduce their abundances and grain
surface reactions on icy mantles are often invoked. However, it is
not clear what mechanisms are responsible for the release of icy
mantles in cold cores; it is unlikely that non-thermal desorption
alone would suffice and the presence of shocks inducing explosive
injection of grain mantles has been suggested in the past \citep[e.g.][]{S06}.
Our model provides a novel mechanism for the production of large complex 
molecules in cold dark clouds. 

\item 
Methanol (CH$_3$OH) and formaldehyde (H$_2$CO) are found in ices in quiescent 
regions. In our model these species are synthesised in the gas phase following
the ice mantle explosions, are ejected and subsequently frozen onto the 
grains in the next cycle. 
This is different from the conventional picture in which they are made by the
hydrogenation of CO in the ice. Although we allow for the partial conversion 
of CO to CO$_2$ we have deliberately suppressed the solid-state 
conversion of CO to CH$_3$OH in our models, so as to clearly separate the 
significance of our mechanism from the conventional channel. 
We find that both processes may be effective.

\item 
The composition of the ice mantles and the nature of gas-grain 
interactions differ from previous models.
We have postulated an alternative desorption mechanism which should be 
considered alongside other desorption processes.
Our model predicts that the ices will probably be more chemically complex 
than previously recognised. 
Moreover, the chemical processing is cumulative, so that we expect the 
the chemical complexity of ices in molecular clouds to grow in time.

\item 
The presence of trace metal elements, such as Na, in the hot sublimated
gas together with OH radicals may lead to the formation of NaOH.
NaOH has a large dipole moment, a rotational constant of $B_0\sim 12.57$\,GHz 
and an observationally detectable rotational spectrum \citep{PT73}.

\item 
The chemical network that we employ in the high density phase (Phase II) is 
highly speculative, probably incomplete, and somewhat pessimistic in that it
only involves reactions with H$_2$O as a {\em passive} third body; i.e. one 
that stabilizes associating species, but does not chemically react with them.
Our results should therefore be considered as lower limits to the large 
molecule formation efficiencies.
We can speculate, for example, that the abundant species CO (which is 
technically a radical) could be involved as an {\em active} third body in 
which it chemically combines with the other reactants. If, for example, these
were NH$_2$ and CH$_2$OH radicals then it may be possible to form a large 
molecule, such as glycine (NH$_2$CH$_2$COOH), in a single step process.

\item
The model may provide a mechanism for the formation of larger molecules, of
biochemical importance, in molecular clouds. If we consider C$_2$H$_4$O$_2$, 
then three isomers of this molecule are detected in hot cores; acetic acid 
(CH$_3$COOH), methyl formate (HCOOCH$_3$) and glycolaldehyde (CH$_2$OHCHO). 
No detections of these species have been made in molecular clouds. 
However, in hot cores, acetic acid and glycolaldehyde are only detected rarely 
(such as in Sgr B2(N) and W51), whilst methyl formate is more abundant and is 
observed in many sources. Thus, for example, the relative abundances
of (acetic acid):(glycolaldehyde):(methyl formate) are 
$\sim$1:4:26 in Sgr B2(N). 
This is consistent with the observation that interstellar molecules with a 
C-O-C backbone structure are preferred over those with a C-C-O structure
\citep{HLJ00}.

Our model only specifically identifies glycolaldehyde as a reaction product.
It is possible that methyl formate could be formed as a result of a three-body
reaction involving the radicals CH$_3$O and HCO.
However, even if this assumption is made, then the model predicts methyl
formate abundances that are less than those of glycolaldehyde. There are 
three possible explanations for this apparent discrepancy: 
(i) The physical and chemical conditions in hot cores are very different
to molecular clouds, so that molecular ratios determined from hot core 
observations may not apply,  
(ii) as is highly 
likely to be the case, our Phase II reaction network is neither fully 
accurate, nor complete and/or (iii) alternative mechanisms may be operating 
which produce the methyl formate. 
Indeed a variety of mechanisms have been proposed for methyl formate formation,
including the solid-state reaction of methanol and CO in cosmic-ray irradiated 
ices \citep{MP10,O11}, the solid-state reaction of formic acid and methanol 
(Hollis et al. 2000) and gas-phase reaction of methanol and formaldehyde in 
hot cores \citep{CHH93}. 

Looking at the structure of these three isomers we see that methyl formate, 
glycolaldehde and acetic acid have a central O-atom (ether group), methylene 
group and carbonyl group respectively. As we do not have atomic oxygen in our 
Phase II chemistry it is therefore perhaps not surprising that methyl formate
is not produced in our model.

If we go on to speculate about the possible formation of larger bio-molecules 
then we should note that methyl formate is less important that the other two 
(C-C-O structure) isomers; glycolaldehyde is the simplest sugar, whilst acetic 
acid is only an NH$_2$ group away from the simplest amino acid, 
glycine (NH$_2$CH$_2$COOH).
Indeed, it is highly likely that the main synthesis channels for glycine 
involve acetic acid so that it may be something of an essential pre-cursor to 
bio-molecule formation \citep{S06}. 

Whilst acetic acid is not specifically included in our reaction scheme, it is a
perfectly feasible product of the reactants that are present in the high 
density phase (Phase II) of our model - indeed more so than methyl formate.
It is also worth noting that our model does produce formic acid (HCOOH) with 
typical fractional abundances of $\sim 10^{-10}-5\times 10^{-9}$. 
Formic acid is also 
believed to be a significant species in glycine synthesis \citep{S06}.
An alternative mechanism for the formation of complex organic molecules was
proposed by \citet{S01} in which radicals are created in ice mantles by UV
photolysis. 
In that model, the grains are heated by grain-grain collisions and the radicals 
react with each other {\em in the mantles}. Chemical explosions release the 
complex organics, including glycine and glycolaldehyde into the gas-phase.
There is no reason why the mechanism that we are proposing and that of
\citet{GWWH08} should not be both possible and may indeed complement each 
other.

\end{itemize}

\bibliographystyle{mn2e}

\end{document}